\documentclass[12pt]{article}
\usepackage{float}
\usepackage{wrapfig}
\usepackage[scaled=0.92]{helvet}   

\usepackage[T1]{fontenc}
\usepackage{amsmath}
\usepackage{amsfonts}
\usepackage{pgf}
\usepackage{fancyhdr}
\fancypagestyle{newstyle}{
  \fancyhf{} 
  \fancyfoot[R]{\vspace{0.1in} \small \thepage}
  
  }
\usepackage[top=1in,
bottom=1in,
left=1in,
right=1in]{geometry}

\usepackage{natbib}
\usepackage{hyperref}
\usepackage{epsfig}
\usepackage{url}
\usepackage{bm}

\usepackage{titlesec}
\titleformat*{\section}{\normalsize\bfseries}
\titleformat*{\subsection}{\normalsize\bfseries}

\usepackage{enumitem}
\setlist{nolistsep}

\graphicspath{{figs/}}

\begin{document}

\title{Causal inference for process understanding in Earth sciences}

\author{Adam Massmann\thanks{Corresponding author:
    akm2203@columbia.edu}, Pierre Gentine, Jakob Runge}

\maketitle
\begin{abstract}
  There is growing interest in the study of causal methods in the
  Earth sciences. However, most applications have focused on causal
  discovery, i.e. inferring the causal relationships and causal
  structure from data. This paper instead examines causality through
  the lens of causal {\it inference} and how expert-defined causal
  graphs, a fundamental from causal theory, can be used to clarify
  assumptions, identify tractable problems, and aid interpretation of
  results and their causality in Earth science research. We apply
  causal theory to generic graphs of the Earth system to identify
  where causal inference may be most tractable and useful to address
  problems in Earth science, and avoid potentially incorrect
  conclusions. Specifically, causal inference may be useful when: (1)
  the effect of interest is only causally affected by the observed
  portion of the state space; or: (2) the cause of interest can be
  assumed to be independent of the evolution of the system’s state;
  or: (3) the state space of the system is reconstructable from lagged
  observations of the system. However, we also highlight through
  examples that causal graphs can be used to explicitly define and
  communicate assumptions and hypotheses, and help to structure
  analyses, even if causal inference is ultimately challenging given
  the data availability, limitations and uncertainties.
\end{abstract}

\paragraph{Note:} We will update this manuscript as our understanding
of causality's role in Earth science research evolves. Comments,
feedback, and edits are enthusiastically encouraged, and we will add
acknowledgments and/or coauthors as we receive community
contributions. To edit the manuscript directly (recommended) you can
fork the project's repository and submit a pull request at
\url{https://github.com/massma/causality-earth-science}, or you can
alternatively email us comments, questions, and suggestions and we
will try to incorporate them into the manuscript.

\section{Introduction}

There is growing interest in the study of causal inference methods in
the Earth sciences \citep[e.g.,][]{salvucci2002, ebert-uphoff2012,
  kretschmer2016,Green_2017,barnes-2019,
  samarasinghe2020,runge-causal-timeseries,runge2019inferring,goodwell-causality-2020}. However,
most of this work focuses on causal discovery, or the inference (using
data) of causal structure: i.e., the ``links'' and directions between
variables. In some cases, causal discovery can be used to estimate the
structure of a causal graph and the relationships between variables
when the graph is not known a priori. However in many if not most
Earth system applications, the causal graph is already known based on
physical insights. For instance the impact of El Ni\~{n}o on West
American rainfall is known to be causal and the graph does not need to
be discovered (even though using causal discovery for this problem is
a useful sanity check of the method's ability).

This paper looks at causality through a different, but complementary,
lens to causal discovery and examines how assumed causal graphs \citep{pearl1995causal}, a
fundamental from causal theory, can be used to clarify assumptions,
identify tractable problems, and aid interpretation of results in
Earth science research. Our goal is to distill
\citep[e.g.,][]{olah2017} the basics of the graphical approach to
causality in a way that is relatable for Earth scientists,
\textbf{hopefully motivating more widespread use and adoption of
  causal graphs to organize analyses and communicate
  assumptions}. These tools are relevant now more than ever, as the
abundance of new data and novel analysis methods have inevitably led
to more opaque results and barriers to the communication of
assumptions.

Beyond their usefulness as communication tools, if certain conditions
are met, causal graphs can be used to estimate, from data, the
generalized functional form of relationships between Earth science
variables \citep{pearl2009causality}. Ultimately, deriving generalized
functional relationships is a primary goal of science. While we know
the functional relationships between some variables a priori, there
are many relationships we do not know \citep[e.g., ecosystem scale
water and carbon fluxes;][]{massmann2019, zhou2019arid,
  zhou2019feedback, grossiord2020}, or that we do know but are
computationally intractable to calculate \citep[e.g., clouds and
microphysics at the global scale:][]{randall2003, gentine2018,
  zadra2018, gagne2020emulation}. In these types of applications,
causal graphs give us a path toward new scientific knowledge and
hypothesis testing: generalized functional relationships that were
inaccessible with traditional tools.

The main contribution of this paper is to demonstrate how causal
graphs, a fundamental tool of causal inference discussed in Section
\ref{sec:what-caus-caus}, can be used to communicate assumptions,
organize analyses and hypotheses, and ultimately improve scientific
understanding and reproducibility. We want to emphasize that almost
any study could benefit from inclusion of a causal graph in terms of
communication and clarification of hypotheses, even if in the end the
results cannot be interpreted causally. Causal graphs also encourage
us to think deeply in the initial stages of analysis about hypotheses
and how the system is structured, and can identify infeasible studies
early in the research process before time is spent on analysis,
acquiring data, building/running models, etc. These points require
some background and discussion, so the paper is divided into the
following sections:

\begin{itemize}
\item Section \ref{sec:what-caus-caus}: Introduces and discusses
  causal graphs within the general philosophy of causality and its
  application in Earth science.
\item Section \ref{sec:causal-graphs-pearls}: Using a simple relatable
  example we explain the problem of confounders and how causal graphs
  can be used to isolate the functional mapping between interventions
  on some variable(s) to their effect on other variable(s).
\item Section \ref{sec:causal-graphs-as}: We draw on a real example
  that benefits from inclusion of a causal graph, in terms of
  communicating assumptions, and organizing and justifying analyses.
\item Section \ref{sec:necess-cond-caus}: We turn to more generic
  examples of graphs that are generally consistent with a wide variety
  of systems in Earth science, to highlight some of the difficulties
  we confront when using causal inference in Earth science and how we
  may be able to overcome these challenges.
\end{itemize}

Throughout the discussion key terms will be emphasized in italics.

\section{The graphical
  perspective to causality and its usefulness in the Earth
  sciences}\label{sec:what-caus-caus}

\begin{wrapfigure}{L}{0.35\textwidth}
  \includegraphics[]{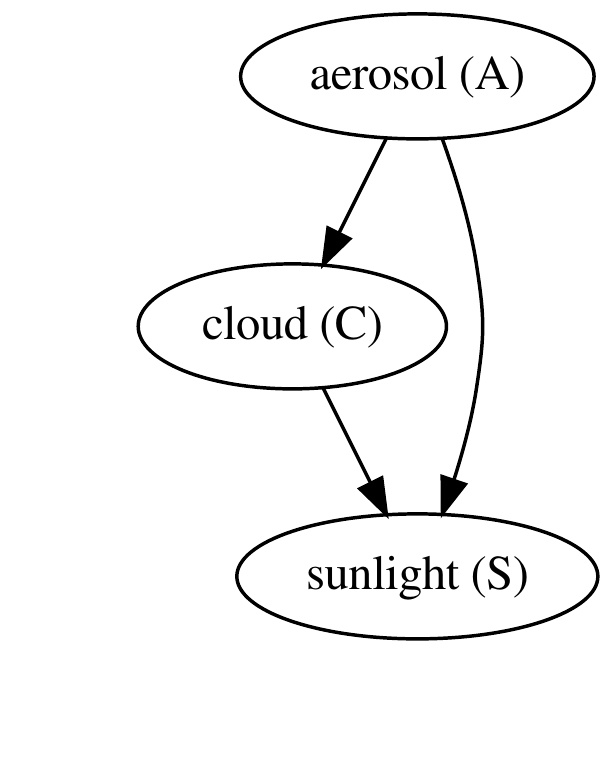}
  \caption{A toy graph example to demonstrate basic causal theory,
    involving cloud (C), aerosol (A), and surface solar radiation
    (S).}
  \label{fig:toy}
\end{wrapfigure}

While there are many different definitions and interpretations of
``causality'', for this manuscript we view causality through the lens
of causal graphs, as introduced in \citet{pearl1995causal} and
discussed more extensively in \citet{pearl2009causality}. We take this
perspective because we believe causal graphs are useful in Earth
science, rather than because of any particular philosophical argument
for causal graphs as the ``true'' representation of causality.

Causal graphs are Directed Acyclic Graphs (\emph{DAG}s) that encode
our assumptions about the causal dependencies of a system. To make a
causal graph, a domain expert simply draws directed edges
(i.e. arrows) from variables that are causes to effects. In other
words, to make a causal graph you draw a diagram summarizing the
assumed causal links and directions between variables (e.g., Figure
\ref{fig:toy}). Causal graphs are useful tools because they can be
drawn by domain experts with no required knowledge of maths or
probability, but they also represent formal mathematical
objects. Specifically, underlying each causal graph are a set of
equations called \emph{structural causal models}: each node
corresponds to a generating function for that variable, and the inputs
to that function are the node's parents. Parents are the other nodes
in the graph that point to a node of interest (e.g., in the most
simple graph $X \to Y$; $X$ is a parent of $Y$). So in reality,
drawing arrows from ``causes'' to ``effects'' is synonymous with
drawing arrows from function inputs to generating functions.

In this way, drawing a causal graph is another way to visualize and
reason about a complicated system of equations, which is a very useful
tool for the Earth scientist: we deal with complicated interacting
systems of equations and welcome tools that help us understand and
reason about their collective behavior. In some cases we may know a
priori (from physics) the equations for a given function in a causal
graph. However, in practice we often either do not know all of the
functions a priori \citep[e.g., plant stomata response to
VPD;][]{massmann2019, zhou2019arid, zhou2019feedback, grossiord2020},
or some functions are computationally intractable to compute
\citep[e.g., turbulence, moist convection, and cloud microphysics in
large scale models;][]{zadra2018,gentine2018}. In these scenarios the
benefits of causal graphs are fully realized: based on the causal
graph we can calculate from data, using the \textit{do-}calculus
\citep{pearl-1994-do-calculus}, the response of target variables
(i.e. effects) to \textit{interventions} on any other variables in the
graph (i.e. causes). When combined with statistical modeling (i.e.,
regression), one can estimate the functional relationship between
interventions on causes and effects
\citep{pearl2009,shalizi2013,shi2019adapting,mao2020generative}.

By viewing causal graphs through this pragmatic lens of calculating
the functional form of relationships that we do not know a priori, we
simultaneously identify causal graphs' value for Earth scientists
while also side stepping philosophical arguments about the meaning of
causality. Causal graphs are pragmatic because in the Earth sciences
we often need to estimate how the system responds to
\emph{interventions} (prescribed changes to variables of interest, or
``causes''). For example, sub-grid physical parameterizations in Earth
system models (e.g., turbulence) require estimates of the time
tendencies' response to \emph{interventions} on the large scale state
and environment. We also may desire to calculate experiments: for
example how changing land cover from forest to grasslands affects (the
statistics of) surface temperature. \textit{Do-}calculus is a method
to calculate this response to interventions without relying on
approximate numerical models or real world experimentation, which can
be infeasible or unethical \citep[as is the case for geoengineering;
e.g., unilateral decisions to seed the oceans with iron, or spray
aerosols in the atmosphere,][]{hamilton2013no}. While we want to
maintain this emphasis on causality as a method for calculating the
generalized response to intervention (possibly using regression to
calculate the functional form of that response), for consistency with
the causal literature we will call the response variables ``effects'',
and the intervened-upon variables ``causes''.

For some, it may not be clear how the functional response to
\emph{interventions} is different from naive regression between
observed variables. We will demonstrate in Section
\ref{sec:causal-graphs-pearls} how uninformed regression is just a
functional mapping of associations between variables, and how this
differs from the response to interventions, i.e. a causal
mechanism. This is the problem that \textit{do-}calculus solves: it
identifies which data are needed and how we can use those data to
calculate the response to \emph{interventions}, rather than just
associations that may be attributable to other processes
entirely. Because we know the response to the intervention is
attributable to the intervention and not other processes, we have
greater confidence in the generalizability of the response to
interventions.  This generalizability of the response to interventions
makes \textit{do-}calculus especially relevant for scientists and
engineers. Working through an example will clarify some of these
claims.

\section{A Toy Example: the problem of confounding and the necessity
  of \textit{do-}calculus for calculating interventions}
\label{sec:causal-graphs-pearls}

To demonstrate the problem of \textit{confounding} and the necessity
of causal graphs/\textit{do-}calculus, we use a simple toy
example involving clouds, aerosols, and surface solar
radiation/sunlight. As shown in Figure \ref{fig:toy}, the causal
graph consists of:

\begin{enumerate}
\item An edge from aerosols to clouds because aerosols serve as cloud
  condensation nuclei or ice nucleating particles, which affect the probability of water vapor conversion to cloud (condensates).
\item An edge from aerosols to surface solar radiation, because
  aerosols can reflect sunlight back to space and reduce sunlight at
  the surface.
\item An edge from clouds to sunlight, because clouds also reflect
  sunlight back to space and can reduce sunlight at the surface.
\end{enumerate}

Causal graphs encode our assumptions about how the system behaves, and
the nodes and edges that are \textit{missing} from the graph often
represent strong assumptions about the lack of functional
dependence. For example, in the cloud-aerosol-sunlight example, clouds
also affect aerosols; e.g., by increasing the likelihood that aerosol
will be scavenged from the atmosphere during precipitation
\citep[e.g.,][]{radke-scavenge-1980, jurado2008,
  blanco-alegre2018}. By not including an edge from cloud to aerosol,
we are making the assumption that we are neglecting the effect of
clouds on aerosols, and also preventing the graph from containing any
cycles (a path from a variable to itself) which is a requirement of
the theory: graphs must by acyclic. This acyclic requirement may raise
concerns for the reader; many problems in Earth science contain
feedbacks that introduce cycles. However, any feedback can be
represented as an acyclic graph by explicitly resolving the time
evolution of the feedback in the graph (Section
\ref{sec:necess-cond-caus} contains examples of such graphs).
Considering this example is intended to be pedagogical for introducing
causal theory to the readers, we will continue with the graph as drawn
in Figure \ref{fig:toy} (we refer the reader to \cite{gryspeerdt-2019}
for a more realistic treatment of aerosols and clouds).

Even though mathematical reasoning is not required to construct a
causal graph, the resulting graph encodes specific causal meaning
based on qualitative physical understanding of the system. Implicitly,
the graph corresponds to a set of underlying functions, called a
structural causal model, for each variable:

\begin{align}
  \label{eq:2}
  aerosol &\leftarrow f_{aerosol} (U_{aerosol}) \\
  cloud &\leftarrow f_{cloud} (aerosol, U_{cloud})\\
  sunlight &\leftarrow f_{sunlight} (aerosol, cloud, U_{sunlight})
\end{align}

where $U$ are random variables due to all the factors not represented
explicitly in the causal graph, and $f$ are deterministic functions
that generate each variable in the graph from their parents and
corresponding $U$.

The presence of the random variables $U$ introduces a third meaning to
the causal graph: they induce a factorization of the joint
distribution between variables into conditional and marginal factors:
\begin{equation}
  P(A, C, S) = P(S \, | \,C, A) \, P(C \, | \, A) \, P(A),
\end{equation}
where $A$ represents aerosol, $C$ represents cloud, $S$ represents
surface solar radiation, and $P($C$ | $A$)$ denotes conditional
probability of $C$ given $A$ (Appendix \ref{prob-theory} describes the notation used in
this paper and a brief introduction to probability theory for
unfamiliar readers). The inclusion of randomness in causal graphs is a
key tool: by positing a causal graph, we are not stating that the
variables in the graph are the only processes in the system nor that the relationships are deterministic. Instead,
we are stating that all other processes not included in the graph
induce variations in the graph's variables that are \emph{independent} of
each other (e.g., all $U_{\cdot}$ in Equation (\ref{eq:2}) are
independent). For example, sources of aerosol variability not
considered in Figure \ref{fig:toy} include anthropogenic aerosol
emission, the biosphere, fires, volcanoes,
etc. \citep[e.g.,][]{Boucher2015}. For cloud, this includes synoptic
forcing or atmospheric humidity,
etc. \citep[e.g.,][]{wallace2006atmospheric}. For radiation, this
includes variability of top of atmosphere radiation,
etc. \citep[e.g.,][]{hartmann2015global}. Figure \ref{fig:toy} states
that all these external, or \textit{exogenous}, sources of variability
are independent of each other \citep[in very technical terms, this means the
graph is ``\textit{Markovian},''][]{pearl2009causality}).

We can now apply causal inference theory
\citep[e.g.,][]{pearl1995causal,tian2002general,shpitser2006} to the
assumptions encoded in our causal graph to identify which
distributions must be estimated from data in order to calculate the
response of effect(s) (e.g. of sunlight) to an experimental
intervention on the cause(s) (e.g. presence or absence of a
cloud). The goal of causal inference is to derive the response to the
intervention in terms of only observed distributions. This process of
identifying the necessary observed distributions is formally termed
\emph{causal identification} \citep[][, Ch. 3]{pearl2009causality}. If
a causal effect is not identifiable (\emph{un}-identifiable), for
example if calculating a causal effect requires distributions of
variables that we do not observe, then we cannot use causal inference
to calculate a causal effect, even with an infinite sample of
data.

A necessary condition for \emph{unidentifiability} is the presence of
an unblocked \emph{backdoor path} from cause to effect \citep[][,
Ch. 3]{pearl2009causality}. Backdoor paths are any paths going through
parents of the cause to the effect. We can block these paths by
selectively observing variables such that no information passes
through them \citep{geiger-d-sep}. If observations are not available
for the variables required to block the path, the path will be
\emph{unblocked}. However, if we can observe variables along the
backdoor paths such that all backdoor paths are blocked, then we have
satisfied the \emph{back-door criterion} \citep{pearl2009} and we can
calculate unbiased causal effects from data.

Understanding backdoor paths and the backdoor criterion is helped by
an example. Returning to our toy example (Figure \ref{fig:toy}), we
attempt to calculate the causal effect of clouds on sunlight. In other
words, we want to isolate the variability of sunlight due to the
causal link from cloud to sunlight (Figure \ref{fig:toy}). However,
aerosols affect both cloud and sunlight (i.e., there is a backdoor
path from cloud to aerosol to sunlight), so if we naively calculate a
"causal" effect using correlations between sunlight and cloud, we obtain
a biased estimate. To demonstrate this, consider simulated cloud,
aerosol, and sunlight data from a set of underlying equations
consistent with Figure \ref{fig:toy} and Equation (\ref{eq:2}):

\begin{align}
  \begin{split}
    \text{aerosol} =& \; U_{aerosol}; \; U_{aerosol} \sim
    \text{uniform (0, 1]}\\ \text{cloud} =& \; \text{Cloudy if } U_{cloud} +
    \text{aerosol} > 1; \; U_{cloud} \sim \text{uniform (0, 1]}\\ \text{sunlight}
    =& \begin{cases} \text{Cloudy} &: 0.6 \cdot \text{downwelling clear
        sky radiation} \\ \text{Clear} &: \text{downwelling clear sky
        radiation}
    \end{cases}
    \label{eq:1}
  \end{split}
\end{align}

where:

\begin{equation*} \text{downwelling clear sky radiation} =
  U_{sunlight} \cdot (1 - aerosol); \; U_{sunlight} \sim
  \text{Normal(340 W m$^{-2}$, 30 \, W m$^{-2}$)}
\end{equation*}

\begin{figure} \centering \includegraphics[]{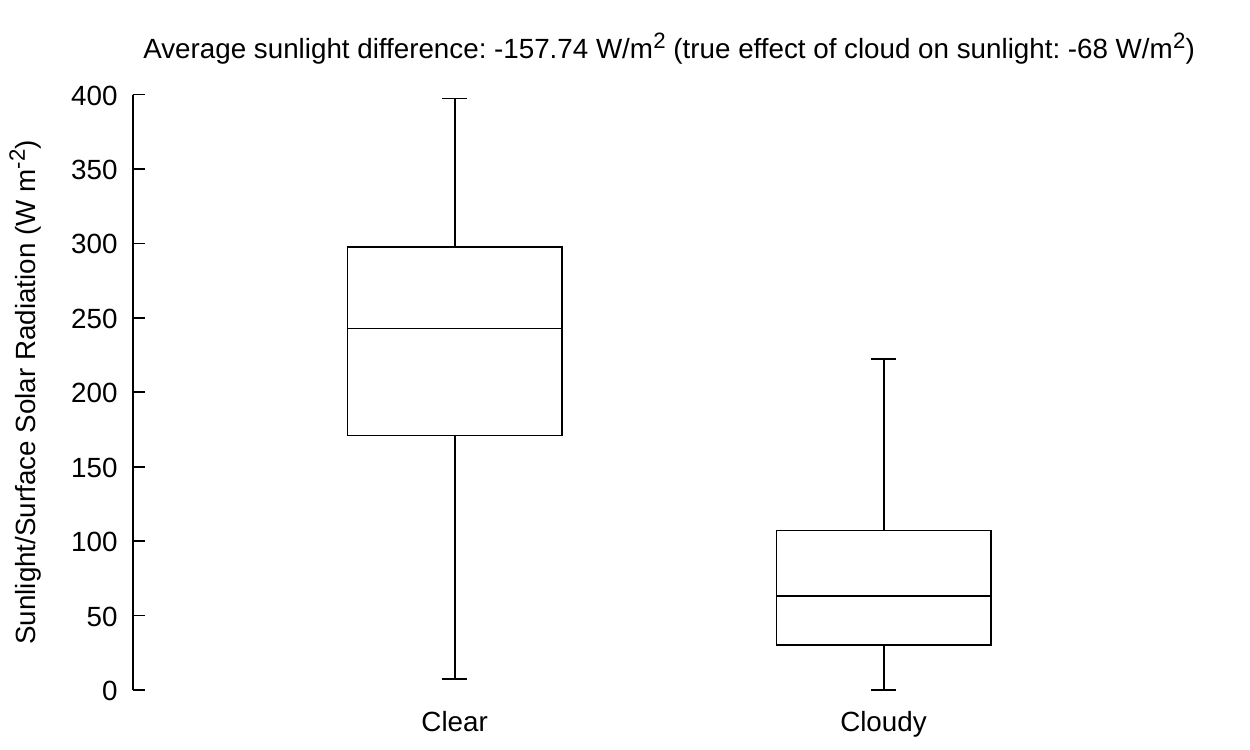}
  \caption{A naive approach to estimating the ``effect'' of clouds on
    sunlight: bin observations by cloudy and clear day, and compare the
    values of sunlight. This approach yields an average difference of
    157.74 W m$^{-2}$ between cloudy and clear days, and is a large
    overestimation of the true causal effect of clouds on sunlight (-68.0
    W m$^{-2}$) in these synthetic data.}
  \label{fig:naive-cloud-sunlight}
\end{figure}

Now, consider not knowing the underlying generative processes, but
instead just passively observing cloud and sunlight. If one were
interested in calculating the effect of cloud on sunlight, and aerosol
data were not available or one were not aware that aerosol could have
an impact on clouds and aerosol, one direct but incorrect approach
would be to bin the data by cloudy and clear conditions and compare
the amount of sunlight between cloudy and clear observations (Figure
\ref{fig:naive-cloud-sunlight}). This approach would suggest that
clouds reduce sunlight by, on average, 158 W m$^{-2}$; this is is a
strong overestimation of the true average effect of clouds (-68 W
m$^{-2}$), derived from Equation (\ref{eq:1}). This overestimation is
due to aerosol-induced co-variability between cloud and sunlight that
is unrelated to the causal link from cloud to sunlight. However if
aerosols were constant (e.g. observed or not varying), any
co-variability between cloud and sunlight would be attributable to the
causal edge between cloud and sunlight (Figure \ref{fig:toy}). In
other words, conditional on aerosol, all co-variability between cloud
and sunlight is only due to the causal effect of cloud on sunlight.
We can mathematically encode this requirement that we must condition
on aerosol to isolate the causal effect of cloud on radiation, and
doing so derives the causal effect of cloud on sunlight by
satisfying the backdoor criterion with \textit{adjustment} on aerosol:

\begin{equation} P(S | do(C = c)) = \int_{a} P(S \, | \, C = c, A=a)
  \, P(A=a) \; da,
  \label{eq:3}
\end{equation} where the \textit{do}-expression ($P(S \, | \, do(C\, = \,c))$) represents the probability of sunlight
if we did an experiment where we intervened and set cloud to a value
of our choosing (in this case $c$, which could be ``True'' for the
presence of a cloud, or ``False'' for no cloud). In the case that observations of aerosols are not available, our causal effect is not identifiable and
we cannot generally use causal inference without further assumptions,
no matter how large the sample size is of our data.

Causal graphs are therefore powerful analysis tools: after encoding
our domain knowledge in a causal graph, we can analyze the available
observations to determine whether a causal calculation is possible,
\textit{without needing to collect, download, or manipulate any
  data}. For more complicated graphs, causal identification can be
automated \citep[][ and \url{http://www.dagitty.net/},
\url{https://causalfusion.net}]{tian2002general,shpitser2006,huang2006identifiability,Bareinboim7345,
  textor2017}. We later use this theory to assess assumptions that
lead to tractable causal analyses for generic Earth science scenarios
(Section \ref{sec:necess-cond-caus}).

Once we have established that a causal effect is identifiable from
data, we must estimate the required observational distributions
(Equation (\ref{eq:3})) from data. Often it may be more
computationally tractable to calculate an average causal effect,
rather than the full causal distribution $P(S | do(C=c))$, which might be difficult to estimate. Returning
to our toy example (Figure \ref{fig:toy}), the average effect is
defined as:

\begin{equation} \mathbb{E}(S | do(C = c)) = \int_{s} s \, P(S = s |
  do(C=c)) \, ds,
  \label{eq:4}
\end{equation}

where $\mathbb{E}$ is the expected value. Substituting Equation
(\ref{eq:3}) into Equation (\ref{eq:4}), and rearranging gives:

\begin{equation} \mathbb{E}(S | do(C = c)) = \int_{a} P(A=a) \;
  \mathbb{E}(S \, | \, C=c, A=a) \, d a,
  \label{eq:5}
\end{equation}

Where $\mathbb{E}(S \, | \, C=c, A=a)$ is just a regression of
sunlight on cloud and aerosol. Estimating the marginal $P(A)$ is
difficult, but if we assume that our observations are independent and
identically distributed (IID) and we have a large enough sample, we
can use the law of large numbers to approximate Equation (\ref{eq:5})
as \citep{shalizi2013} :

\begin{equation} \mathbb{E}(S | do(C = c)) \approx \frac{1}{n}
  \sum_{i=1}^n \mathbb{E}(S \, | \, C=c, A=a_i).
  \label{eq:6}
\end{equation}

Data or prior knowledge can inform the estimate of
$\mathbb{E}(S | C=c, A=a_i)$, but whatever regression method is used,
it should be checked to ensure it is representative of the data and
there is sufficient signal to noise ratio to robustly estimate the
regression. It is important to note how this estimate is different
from the naive association of $\mathbb{E}(S | C=c)$; in
$\mathbb{E}(S | C=c, A=a_i)$ we are controlling for the impact of
aerosol on sunlight by conditioning on $A$ and including aerosol in
the regression. As we will see, using the \textit{do}-calculus
estimate results in an estimate of the effect of cloud on aerosol that
is very different from the naive association, and close to the true
effect.

\begin{figure} \centering \includegraphics[]{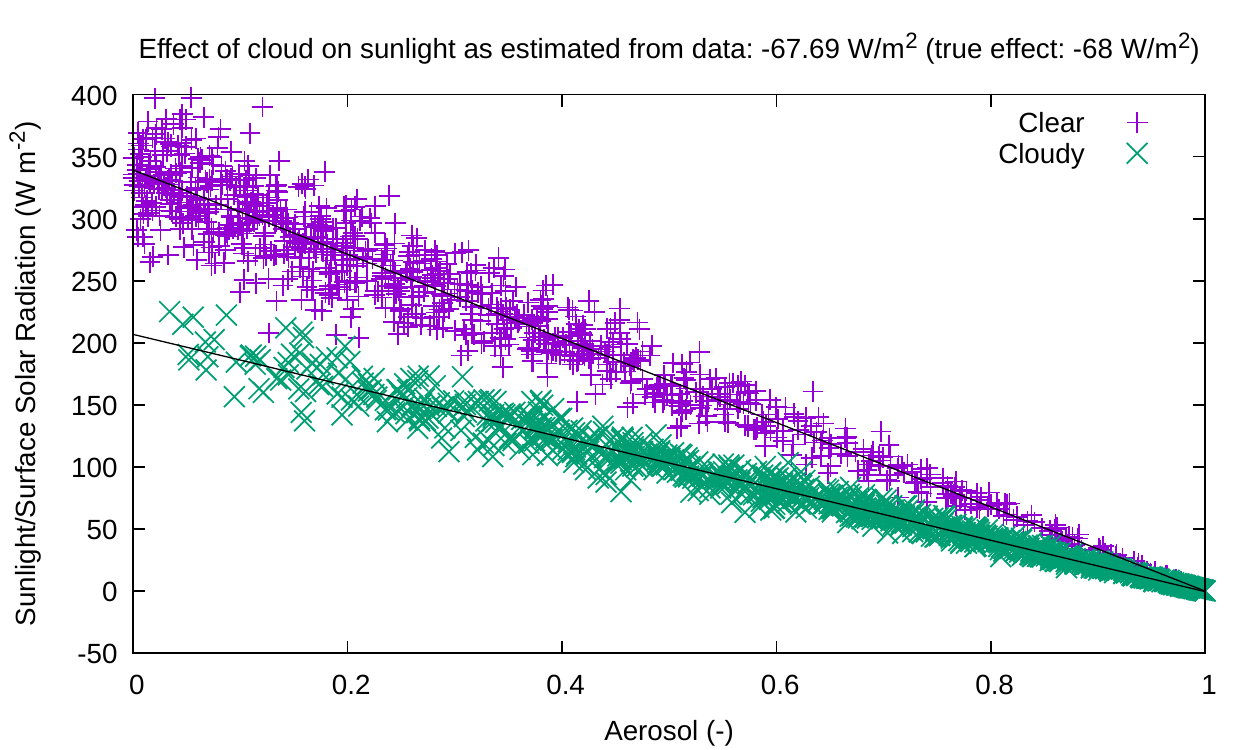}
  \caption{A linear relationship between aerosol and sunlight,
    conditional on cloud. If we use this regression to calculate the
    average causal effect of cloud on sunlight, as in Equation
    (\ref{eq:6}), our result is very close to the true causal effect
    of -68.0 W m$^{-2}$.}
  \label{fig:linear}
\end{figure}

In our simple example, a linear model conditional on cloud is a
suitable choice for the regression function
$\mathbb{E}(S | C=c, A=a_i)$ (Figure \ref{fig:linear}). However, many
problems in Earth science require non-linear approximation methods
like neural networks and/or advanced machine learning methods; for
examples of such machine learning methods, we refer readers to
\citep{bishop2006pattern}.

The causal effect of clouds on sunlight as calculated using Equation
(\ref{eq:6})) (e.g.
$\mathbb{E}(S | do(C = \text{cloudy})) - \mathbb{E}(S | do(C =
\text{clear}))$) is -67.69 W m$^{-2}$, which closely matches the true
causal effect from Equation (\ref{eq:1}) of -68 W m$^{-2}$. This
example demonstrates how causal inference and theory can be used to
calculate unbiased average effects using regression, subject to the
assumptions clearly encoded in the causal graph. Further, causal
inference can be used to justify and communicate assumptions in any
observational analyses employing regression. In the best case, the
causal effect is identifiable from the available observations, and the
regression analysis can be framed as an average causal effect. In the
worst case that identification is not possible from the available
observations, one may present the regression as observed associations
between variables. However, presentation of a causal graph still aids
the reader: the reader can see from the causal graph what the
confounders and unobserved sources of covariability are between the
predictors and the output. In all cases, the presentation of a causal
graph makes explicit the assumptions about the causal dependencies of
the system. Wherever possible, we recommend including causal graphs
with any observation-based analyses.

In summary of the main points of this introduction to causal graphical
models and \textit{do-}calculus:

\begin{itemize}
\item Graphical causal models encode our assumptions about causal
  dependencies in a system (edges are drawn \emph{from} causes
  \emph{to} effects). ``Causal dependencies'' really just refer to
  functional dependencies between inputs (causes) and outputs
  (effect), which are useful in the Earth sciences to reason about
  graphically .
\item In order to calculate an unbiased causal effect from data, we
  must isolate the covariability between cause and effect that is due
  to the directed causal path from cause to effect. The presence of
  non-causal dependencies between the cause and effect can be deduced
  from the causal graph: the presence of an \textit{unblocked backdoor
    path} from the cause to the effect leads to non-causal
  dependencies (and co-variation).
\item The \emph{backdoor criterion} identifies the variables that we
  must condition on in order to block all backdoor paths, remove
  non-causal dependence between the cause and effect, and calculate an
  unbiased causal effect from data.
\item The \emph{average} causal effect can be reliably approximated
  with a regression (Equation (\ref{eq:6})) derived from the backdoor
  criterion. In this scenario, causal theory and graphs identify the
  variables that should be included in the regression in order to
  calculate an unbiased causal effect (however researchers should
  still ensure their choice of regression model is appropriate for the
  data). While not demonstrated by our example, causal theory and
  graphs also identify the variables that \textit{should not} be
  included in the regression \citep{pearl2009}: we can also bias a
  causal effect by including too many variables in a regression .
\item The do-calculus and identification theory provide a flexible
  tool to determine whether an effect is identifiable and, if so,
  which distributions should be estimated from data, while making no
  assumptions about the forms of the underlying functions and
  distributions. However, parametric assumptions can be applied to
  make the calculation of those distributions from data more
  computationally tractable.
\end{itemize}

Here we focused on the \emph{backdoor criterion} to block backdoor
paths. An unblocked backdoor path from the cause to the effect is a
necessary condition for unidentifiability. However, an unblocked
backdoor path from the cause to the effect is not a sufficient
condition for unidentifiability: there are other identification
strategies like the front door criterion \citep[see Section 3.5.2
in][]{pearl2009causality} and instrumental variables \citep[see
Chapter 8 in][]{pearl2009causality} that do not rely on observing
variables along the backdoor path, and can be used in some cases where
observations are not available to satisfy the backdoor criterion (also
see \citet{tian2002general} for more discussion on sufficient
conditions for unidentifiability). These are examples of how the
\textit{do-}calculus admits any strategy that frames the response to
interventions in terms of observed distributions. We focus on the
backdoor criterion because it is the most fundamental and direct
method for adjusting for confounding, the most intuitive for an
introduction to causality, and is the most relevant for the generic
temporal systems present in the Earth sciences (Figure
\ref{fig:generic} in Section \ref{sec:necess-cond-caus}). However,
causal identification through other methods like instrumental
variables and the front door criterion can also be automated; we refer
the reader to \citet{pearl2009causality} for further discussion and
software tools like \url{http://www.dagitty.net/} and
\url{http://www.causalfusion.net} for interactive exploration.

\section{Beyond toys: causal graphs as communicators, organizers, and
  time-savers}\label{sec:causal-graphs-as}

In Section \ref{sec:causal-graphs-pearls} we used a toy example to
demonstrate a causal analysis starting with drawing a graph and ending
with the successful calculation of the average response of sunlight
(the effect) to an intervention on cloud (the cause). However, often
we may not be able to ultimately estimate the causal effects from the
available data. In many cases there are serious challenges due to
unobserved confounding in generic Earth science problems (Section
\ref{sec:necess-cond-caus}), and for other cases we may lack enough
samples, or samples could be too systematically biased, to estimate
the necessary distributions (or regressions; e.g., Equation
(\ref{eq:6})) with sufficient certainty. However, we want to
emphasize that even if calculating a causal effect might in some
cases be impossible, drawing a causal graph at the beginning of an
analysis still offers tremendous benefits in terms of organization and
scientific communication. Investing time to reason about the
functional structure of the problem at the outset can save scientists
time in the long term, forcing us to clarify our thinking early,
expose potential challenges, and identify intractable approaches.

Additionally, once the causal graph is drawn, we can use it as a
communication tool and include it in presentations, papers, and
discussions of our results. Making our assumptions about dependencies
in the system explicit greatly improves the interpretability and
reproducibility of our results. Perhaps our analysis and graph meet
the standards for a causal interpretation, but even if they do not,
the causal graph helps the rest of the community asses the sources of
confounding in the graph that were not controlled for, and understand
if their conceptualization of the graph structure matches the authors'
hypotheses. Often in research there are more assumptions being made
than are communicated, and even when they are communicated, the
assumptions do not always get discussed in a precise way. Including a
causal graph allows the assumptions to be clearly known, and discussed
in a precise and rigorous way \citep[e.g.,][]{hannart-da}.

To support the idea that many analyses would benefit from a causal
graph, we will detail how a past project benefits from a causal
graph. This example also moves beyond the toy example of Section
\ref{sec:causal-graphs-pearls}, and demonstrates causal graphs'
applicability to real problems in Earth science.

\subsection{Causal graphs' utility in a real example}

\begin{figure} \includegraphics[]{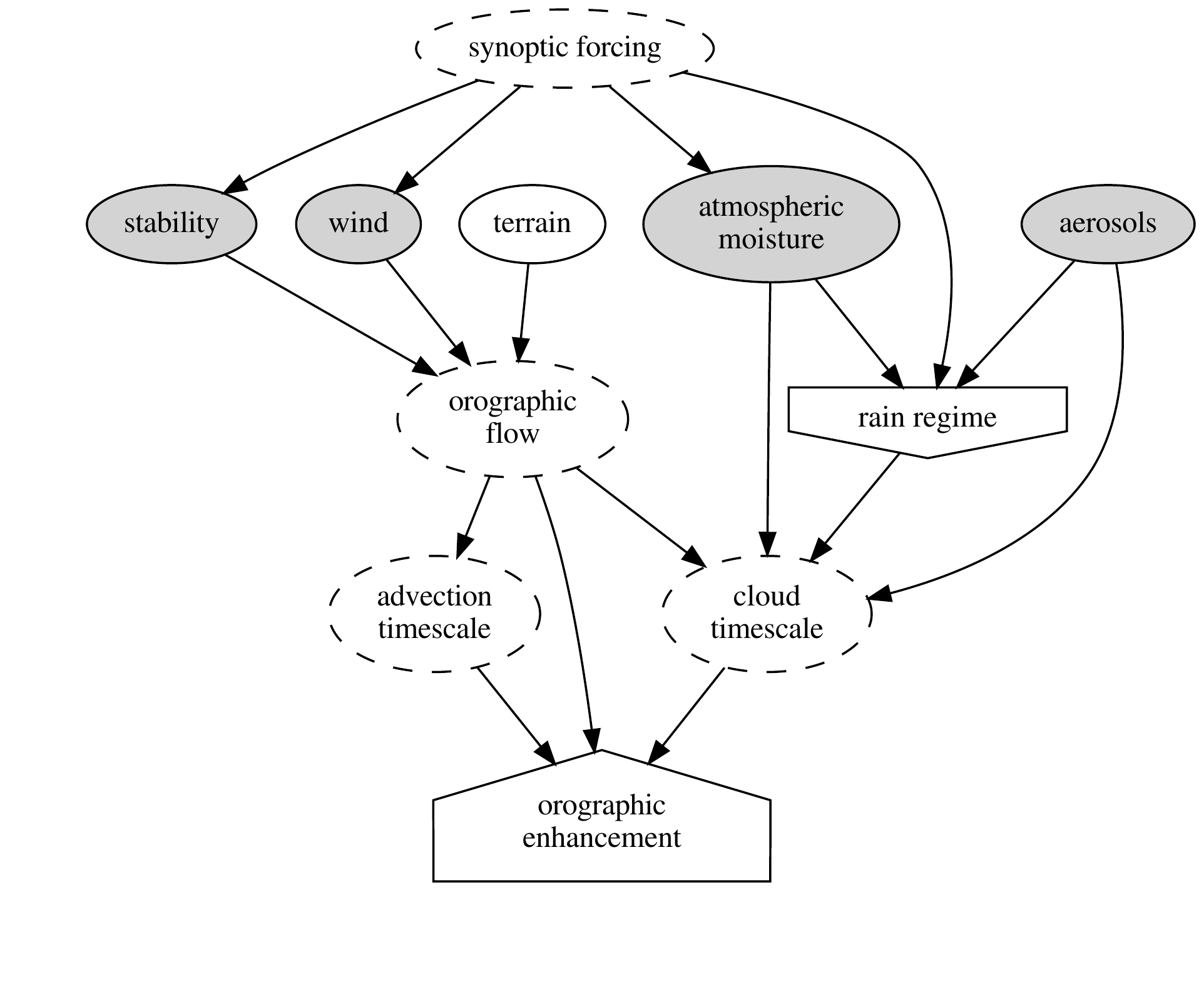}
  \caption{A graph representing steady conditions for orographic
    enhancement during the Chilean Coastal Orographic Precipitation
    Experiment \citep[CCOPE,][]{massmann2017}. The field campaign
    attempted to quantify the effect of ``rain regime'' on
    ``orographic enhancement.''  Observed quantities are represented
    by solid nodes, while unobserved quantities are represented by
    dashed nodes. Variables that must be observed to block all
    backdoor paths are shaded. The effect of rain regime on orographic
    enhancement is identifiable through adjustment on these shaded variables.}
  \label{fig:ccope}
\end{figure}

In \citet{massmann2017}, the lead author of this manuscript
participated in a field campaign designed to study the impact of
microphysical rain regime (specifically the presence of ice from aloft
falling into orographic clouds) on orographic enhancement of
precipitation. This field campaign and analysis benefits from a causal
graph and is a real-world example argument for the more common use of
causal graphs as research tools. Our retrospective causal graph of
orographic enhancement in the Nahuelbuta mountains under steady
conditions clearly communicates our assumptions about the system
(Figure \ref{fig:ccope}).

Many of the variables in the graph, such as ``synoptic forcing'',
``wind'', and ``orographic flow'', are quite general
quantities. Keeping quantities general can lead to more intuitive and
interpretable graphs by limiting the number of details and nodes that
one must consider. However, if logic needs clarifying, the graph can
become more explicit (e.g., differentiating wind into speed,
direction, and spatial distribution, both horizontally and
vertically). We also include unobserved variables explicitly in the
graph, so we can reason about processes' impact on the system even if
they are not observed. As graphs become more complicated, one can
leverage interactive visualization software, or use static graph
abstractions like plates \citep{bishop2006pattern}, which are
particularly well suited to representing repeated structure common to
spatiotemporal systems.

While the exact details of the graph and the assumptions it encodes
are interesting (e.g., ``wind'', ``stability'', and ``atmospheric
moisture'' all refer to upwind conditions, and we assume that these
upwind conditions are the relevant ``boundary condition'' for the
downwind orographic clouds and precipitation), the noteworthy feature
of the graph is that the field campaign's effect of interest, rain
regime on orographic enhancement, is identifiable from the field
campaign's observations. This is subject to the assumptions encoded in
the graph, but those assumptions are explicitly represented and
communicated by the graph. The causal graph helps interpret the field
campaign's results, and in some sense proves that the design of the
field campaign is sound.

Therefore, for field campaigns, causal graphs are particularly useful at the
planning and proposal stage. Such a causal graph could be included in
any field campaign proposal, improving communication about the system
and also rigorously justifying the campaign’s observations as
necessary for calculating the desired effect(s). Even before the
proposal, one could start with a causal graph, and then analyze it to
determine which observations are needed to meet the campaign’s
goals. Building on this idea, one could attach costs associated with
observing each variable in the graph, and automatically determine the
set of observations that minimizes cost while still allowing us to
calculate our effect(s) of interest.

While this is just one example, it demonstrates that causal graphs are
useful beyond toy problems in the Earth system. Additionally, as we
will see in Section \ref{sec:necess-cond-caus}, we can draw quite
general graphs that are representative of many problems in Earth
science. We hope the reader considers drawing a causal graph as a
first step in their next project; they help structure, organize, and
clarify our analysis and its assumptions.

\section{Overcoming unobserved confounding and partial observation of Earth system state}
\label{sec:necess-cond-caus}

So far we have focused on toy (Section
\ref{sec:causal-graphs-pearls}) and specific (Section
\ref{sec:causal-graphs-as}) examples. We now turn our attention to
more general and generic problems in Earth science systems, the common
challenges we may encounter when attempting causal inference, and how
we can overcome these challenges.

Earth science systems and their components are (typically) dynamical
systems evolving through time according to an underlying system state
\citep{lorenz-1963,lorenz1996predictability,majda-state}. This offers
both opportunities and challenges for causal inference. When
constructing causal graphs we may benefit from the temporal ordering
of events \citep{runge2019inferring}: we know that future events can
have no causal effect on the past. We can also use the time dimension
to explicitly resolve feedbacks in the system, and transform cyclic
graphs with feedbacks into directed acyclic graphs (DAGs) required for
causal inference. While handling feedbacks and avoiding cyclic graphs
is a challenge of causal inference in Earth science, resolving the
time dimension is a generic path to overcome this challenge when
observations of sufficient resolution are available.

\begin{figure}
  \centering
  \input{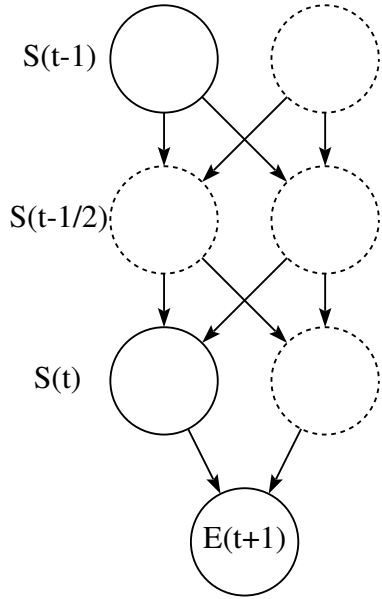}
  \caption{A generic graph of the Earth system state
    sequence. Unobserved nodes are outlined by dashed lines. We only
    observe the state space at certain times (e.g., no observations at
    $S(t-1/2)$), and at times with observations, we only partially
    observe the full state ($S(t)$, $S(t-1)$). In the scenario that we
    are interested in calculating the causal effect of any portion of
    the state space at time $t$ on some effect ($E$) at time $t+1$,
    the causal effect will be confounded by the unobserved portions of
    the state space, and calculating the causal effect is impossible
    (un-identifiable) without additional assumptions.}
  \label{fig:generic}
\end{figure}

However, confounding due to incomplete observation of the system's
state variables also introduces challenges; challenges that are not
unique to this paper's causal lens: incomplete observation of the
system precludes the use of many ``causal discovery'' algorithms as
well (see \citet{runge2019inferring} for a detailed review). Causal
identification and tractable causal inference in Earth science
requires assumptions about the unobserved portions of the state space
that introduce this confounding, and how the unobserved portions of
the state space affect observed variables. Without such assumptions
the unobserved portions of the state space will introduce confounding
for any causal effect of interest (Figure \ref{fig:generic}). For
example, we generally do not observe the state space at every time
(e.g. $S(t-1/2)$, Figure \ref{fig:generic}), and at any given time, we
do not observe the state space at all locations and for all state
variables (e.g. $S(t)$ and $S(t-1)$ in Figure \ref{fig:generic}). In
other words, despite our impressive and growing array of satellite,
remote sensing, and in situ observation systems, we are still very far
from observing every relevant state variable at every location in time
and space.  So, if we are interested in the causal effect of any state
variable at time $t$ on some variable at time $t+1$ (e.g., $E$ in
Figure \ref{fig:generic}), then the causal effect will be confounded
by the unobserved portions of the state space, and calculating a
causal effect will be impossible (un-identifiable) without additional
assumptions.

However, there are assumptions we can make that may be reasonable for
many generic applications, which remove this problem of unobserved
confounding due to partial observation of the state space. We
elaborate upon these assumptions in the following sections, and for
each assumption, we draw a graph, briefly discuss the scenario and
assumptions, present the identification formula and how average
effects can be estimated (for example, using regression to estimate
$\mathbb{E}(\cdot )$), and include some strategies for testing the
assumption(s) with data.

\newpage

\paragraph{Assumption: we observe all state variables that impact our effect(s) of interest}

\begin{figure}[H]
  \centering
  \includegraphics[]{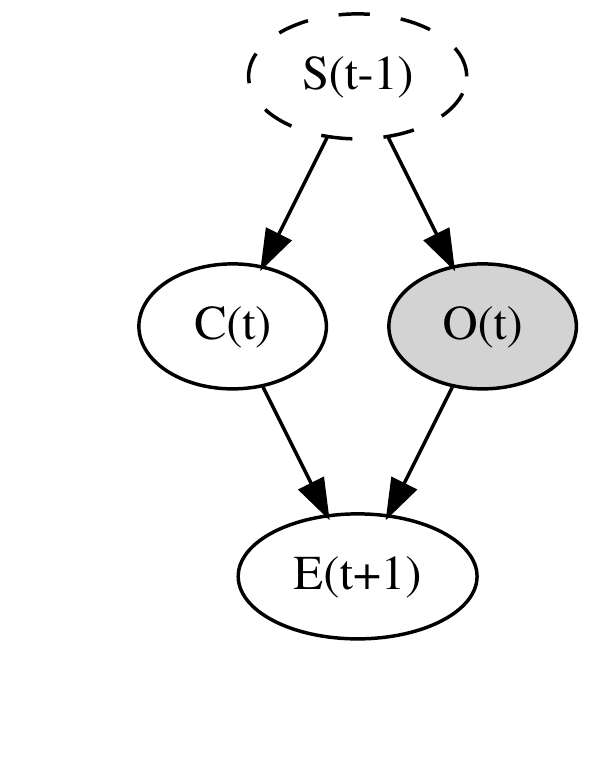}
  \caption{A generic graph for the assumption that we observe all
    state variables that impact our effect(s) of interest. $C(t)$ is
    the cause of interest, $E(t+1)$ is the future effect of interest,
    and $O(t)$ are all observations not including $C(t)$. The full
    state $S(t-1)$ is not observed (dashed node). Observing $O(t)$
    (grey shading) blocks the backdoor path from $C(t)$ to $E(t+1)$.}
  \label{fig:observe-everything}
\end{figure}

\begin{itemize}
\item \textbf{Discussion:} While we may not observe the entire state
  of the system, sometimes it is reasonable to assume that we observe
  the portion of the state space that affects our specific effect of
  interest ($E(t+1)$ in Figure \ref{fig:observe-everything}). In this
  case, we can calculate causal effects by blocking backdoor paths
  with the observed portion of the state space.
\item \textbf{Identification formula:}
  \begin{equation*}
    P(E(t+1) \, | \, do(C(t) = c)) = \int_{o} P(E(t+1) \, | \, C(t) = c,
    O(t) = o) \, P(O(t)=o) \; d o,
  \end{equation*}
\item \textbf{Average causal effect:}
  \begin{equation*}
    \mathbb{E}(E(t+1) \, | \, do(C(t) = c)) \approx \frac{1}{n}
    \sum_{i=1}^n \mathbb{E}(E(t+1) \, | \, C(t)=c, O(t)=o_i),
  \end{equation*}
  where $C(t)$ is the cause of interest, $E(t+1)$ is the effect of
  interest, and $O(t)$ are all observed variables not including $C(t)$.
\item \textbf{Check:} There is no way to check this assumption with
  data. Therefore, this assumption requires strong physical
  justification well supported by the literature. Care is also
  required to insure that there are no interactions between $C(t)$ and
  $O(t)$; e.g., the observations at a given time are truly
  ``simultaneous'' and cannot causally affect each other.
\end{itemize}

\newpage

\paragraph{Assumption: We can reconstruct the state at any given time
  using lagged observations of the system}

\begin{figure}[H]
  \centering
  \includegraphics[]{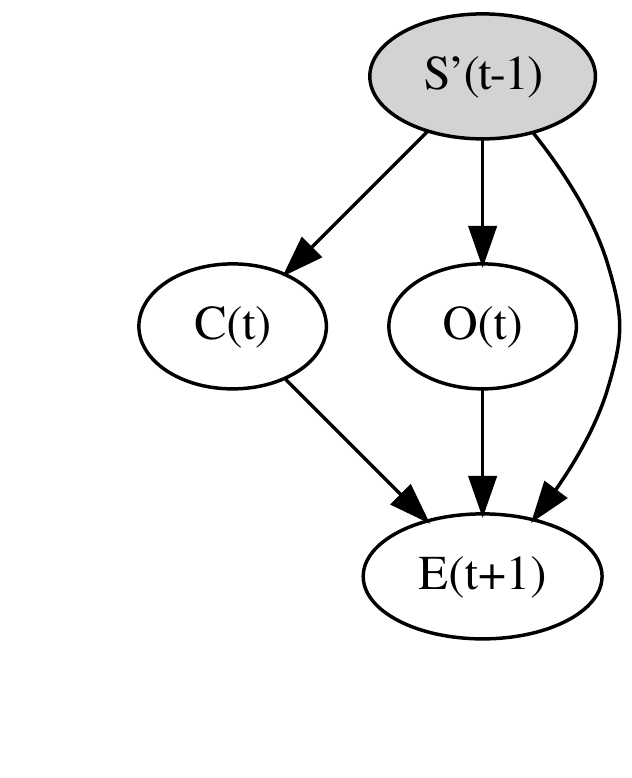}
  \caption{A generic graph for the assumption that we can reconstruct
    the state at any given time using lagged observations of the
    system \citep{takens1981detecting}. $C(t)$ is the cause of
    interest, $E(t+1)$ is the future effect of interest, $O(t)$ are
    all observations not including $C(t)$, and $S'(t-1)$ is the
    reconstructed state. Observing $S'(t-1)$ (grey shading) blocks all
    backdoor paths from $C(t)$ to $E(t+1)$.}
  \label{fig:reconstruction}
\end{figure}

\begin{itemize}
\item \textbf{Discussion:} While we may not observe the entire state
  space, in some cases we may be able to reconstruct the state at any
  given time using lagged observations of the system \citep[see
  Takens' theorem,][]{takens1981detecting}. In this case, we can use
  the reconstructed state to block backdoor paths and examine the
  effect of any observed variable ($C(t)$) on future variables
  ($E(t+1)$ in Figure \ref{fig:reconstruction}). Note that
  additionally controlling for $O(t)$ can also make the effect
  estimate more reliable. More generally, there may be more than one
  suitable set of adjustment co-variates and current research is
  targeted at finding optimal ones that yield the lowest estimation
  error \citep[e.g.,][]{witte2020efficient}.
\item \textbf{Identification formula:}
  \begin{equation*}
    P(E(t+1) \, | \, do(C(t) = c)) = \int_{s} P(E(t+1) \, | \, C(t) = c,
    S'(t-1) = s) \, P(S'(t-1)=s) \; d s.
  \end{equation*}
\item \textbf{Average causal effect:}
  \begin{equation*}
    \mathbb{E}(E(t+1) \, | \, do(C(t) = c)) \approx \frac{1}{n}
    \sum_{i=1}^n \mathbb{E}(E(t+1) \, | \, C(t)=c, S'(t-1)=s_i),
  \end{equation*}
  where $C(t)$ is the cause of interest, $E(t+1)$ is the effect of
  interest, and $S'(t-1)$ is the reconstructed state using lagged
  observations.
\item \textbf{Check:} One check on the success of the state space
  reconstruction is to test whether the observed variables are
  conditionally independent given the reconstructed state
  variable. This approach bears similarity to the deconfounder
  approach introduced by \cite{yixin-2019}, which argues that causal
  effects can be calculated for many problems when we can infer a
  latent variable that renders the (multiple) causes conditionally
  independent given the latent variable.
\end{itemize}

\paragraph{Assumption: the cause of interest is independent of the
  systems' state evolution}

\begin{figure} \includegraphics[]{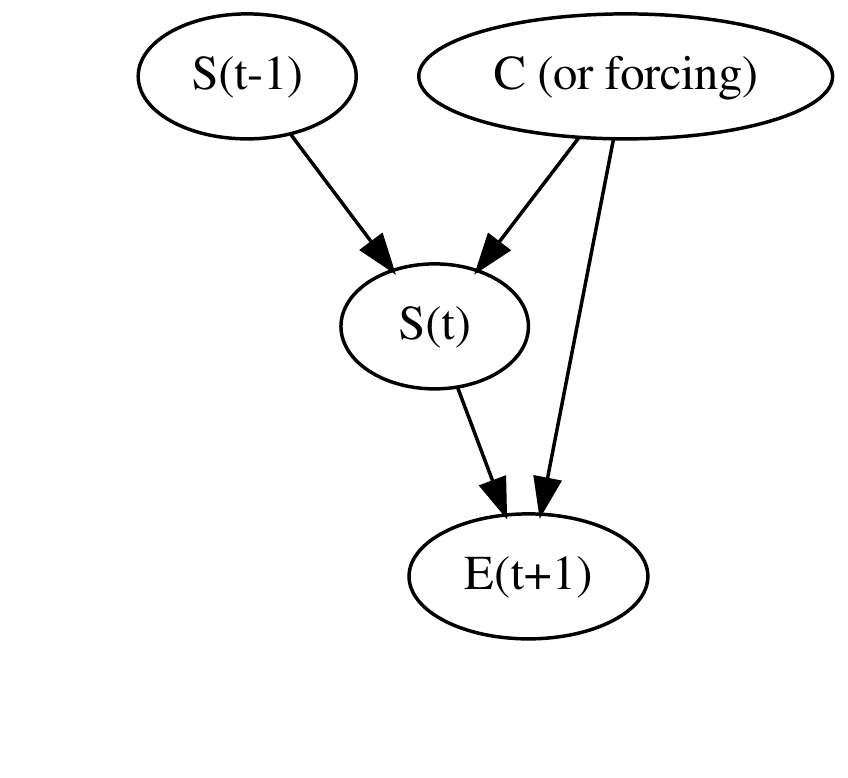}
  \centering
  \caption{A generic graph asserting an assumption that there are
    forcings external to the evolution of the state-space.}
  \label{fig:forcing}
\end{figure}

\begin{itemize}
\item \textbf{Discussion:} In some cases, we may assume that
  the cause of interest is independent of the systems' state
  evolution (Figure \ref{fig:forcing}). While not generally true,
  in some cases a variable may behave independent of the system's
  state while still causally affecting that state. For example, some
  human behavior may be approximated as independent of the climate
  state (e.g., city planning and land use decisions).
\item \textbf{Identification formula:}
  \begin{equation*}
    P(E(t+1) \, | \, do(C = c)) = P(E(t+1) \, | \, C = c)
  \end{equation*}
\item \textbf{Average causal effect:}
  \begin{equation*}
    \mathbb{E}(E(t+1) \, | \, do(C = c)) = \mathbb{E}(E(t+1) \, | \, C=c)
  \end{equation*}
\item \textbf{Check:} To check this assumption, we can test if the
  cause/forcing is independent of the past state. If the cause is
  independent of the past state, we have stronger confidence that the
  assumption holds.
\end{itemize}

While this list of assumptions is certainly not exhaustive, it
presents some approaches that may apply to many scenarios in Earth
science. Also while the provided checks might identify when
assumptions break down, there is no general way to ``validate''
assumptions using data. Physical and science-based justification and
reasoning are ultimately always required.

\section{Conclusions}

In summary, causal graphs and causal inference are powerful tools to
reason about problems in the Earth system, whether in models or observations. Specifically, this review
aimed at showing that:

\begin{itemize}
\item Causal graphs concisely and clearly encode physical assumptions about
  causal/functional dependencies between processes. Including a causal
  graph benefits any observational or modeling analysis, including
  those that use regression.
\item Whether a causal effect can be calculated from data is
  determined by the causal graph. Thus the tractability of a causal
  analysis, or the strength of assumptions necessary to make the
  analysis tractable, is determined and assessed before collecting,
  generating, or manipulating data (which can cost a tremendous amount
  in terms of researchers' time or computational resources). We
  recommend early causal analysis to determine tractability during a
  project's conception, before resources are spent obtaining or
  analyzing data.
\item Calculated causal effects measure the response of target
  variables (i.e. effects) to \textit{interventions} on other
  variables in the system (i.e. causes). With statistical modeling
  (i.e., regression) one can estimate the functional relationship
  between interventions on causes and effects.  These functional
  relationships \textit{generalize} because they map
  \textit{interventions} onto the response: unlike observed
  associations (e.g., naive regression), we know the response is
  attributable to the function's input, and not to some other process in
  the system (e.g., an unobserved common cause). This causal approach opens
  up a new path to calculate generalized functional relationships when
  we either do not know the functional form a priori, or it is too
  computationally intractable to calculate from models.
\item Because the Earth system and its constituents are dynamical
  systems evolving through time, we can construct broadly applicable,
  generic Earth system causal graphs. We can use these graphs to
  calculate generalized functional relationships between processes of
  interest, which would not be possible with associations, correlations
  or simple regressions. However, causal inference in the Earth
  sciences also presents challenges as we only partially observe the
  state space of the system.
\item These challenges can be alleviated by applying causal theory to
  generic causal graphs of the Earth system and identifying the
  assumptions that allow for causal inference from data (Section
  \ref{sec:necess-cond-caus}).
\end{itemize}

Here we focus on the fundamentals of calculating causal effects from
data. However, causal inference is a thriving, active area of
research, and there are many other causal inference techniques and
abstractions that could benefit the Earth system research
community. For example, there are techniques for representing
variables observed under selection bias in the causal graph and
analyzing whether a causal effect can be calculated (i.e. identified)
given that selection bias
\citep[e.g.,][]{bareinboim2014recovering,correa2018generalized}. Selection
bias, defined as a preferential sampling of data according to some
underlying mechanism, is very relevant in Earth sciences. For example,
satellite observations are almost always collected under selection
bias (e.g. clouds obscure surface data, satellites sample at certain
local times of the day which is connected to top of atmosphere solar
forcing, or the sensors themselves could have a bias). Additionally,
transportability
\citep[e.g.,][]{bareinboim2012transportability,Bareinboim7345,lee2019general}
identifies whether one can calculate a causal effect in a passively
observed system called the ``target domain'', by merging experiments
from other systems, called ``source domains'', that may differ from
the target domain. A potential application for transportability in
Earth sciences would be to merge numerical model experiments (e.g.,
Earth system models) and formally transport their results to the real
world. In this case, numerical models are the source domains that
differ from the target domain (``real world'') due to approximations
and different resolutions.

However, because these developments in causal theory are relatively
new, applied domains have yet to establish these recent theoretical
developments' utility for applied analysis. While we encourage applied
scientists to explore how these developments may apply to their
domains, we recognize that many scientists prefer tools with
established utility. To that end, we believe that the use of causal
graphs to organize and structure analyses is mature and directly
applicable to many projects, and can serve as a gateway to applying
these more recent causal developments. We hope that drawing and
including causal graphs in Earth science research becomes more common
in our field, and that this manuscript provides some of the necessary
foundation for readers to attempt using causal graphs in their future
research.

\paragraph{Acknowledgments} The authors want to thank Elias
Bareinboim, Beth Tellman, James Doss-Gollin, David Farnham, and Masa
Haraguchi for thoughtful feedback and comments that greatly improved
an earlier version of this manuscript.

\bibliography{references.bib}

\appendix
\section{Basic probability and syntax}
\label{prob-theory}

In this paper we use capital letters to represent random variables
(e.g., ``$X$''). For example, $P(X)$ is the marginal probability distribution
of a random variable $X$. $P(X)$ is a function of one variable that
outputs a probability (or density, in the case of continuous
variables) given a specific value for $X$. We represent specific
values that a random variable can take with lowercase letters (e.g.,
$x$ in the case of $X$). $P(X)$ is shorthand; a more descriptive but
less concise way to write $P(X)$ is $P(X=x)$ which represents the fact
that $P(X)$ is a function of a specific value of $X$, represented by
$x$. We use both notations, and $P(X)$ has the same meaning as
$P(X=x)$.

For the unfamiliar reader, there are a few basic rules and definitions
in probability that provide relatively complete foundations for
building deeper understanding of probability. These are the
\textbf{sum rule}:

\begin{equation} P(X=x) = \sum_Y P(X=x,\, Y=y)
  \label{eq:sum}
\end{equation}

and the \textbf{product rule}:

\begin{equation} P(X=x, \, Y=y) = P(X = x \, | \, Y=y ) P(Y=y) = P(Y =
  y \, | \, X=x ) P(X=x)
  \label{eq:product}
\end{equation}

The \textit{joint probability distribution} ($P(X=x,Y=y)$) is the
probability that the random variable $X$ equals some value $x$
\emph{and} the random variable $Y$ equals $y$. The joint distribution
is a function of two variables, $x$ and $y$ which are values in the
domains of the random variables $X$ and $Y$ respectively. The
\textit{conditional probability distribution} ($p(X = x \, | \, Y=y
)$) is also a function of two variables $x$ and $y$, but it is the
probability of observing $X$ equal to $x$, given that we have observed
$Y$ equal to $y$. In other words, if we filter our domain to only
values where $Y=y$, then $p(X = x \, | \, Y=y )$ is the probability of
observing $X=x$ in this sub-domain where $Y=y$. The \textit{marginal
  probability distribution} ($P(Y=y)$) is just the probability that $Y$
equals some value $y$, and is a function of only $y$. We can calculate
the marginal probability from the joint distribution by summing over
all possible values values of the other random variables in the joint
(the ``sum rule'' - Equation (\ref{eq:sum})). Additionally, the joint
distribution can factorize into a product of conditional and marginal
distributions (``the product rule'' - Equation
(\ref{eq:product})). These two simple rules can be used to build much
of the theory and applications of probability theory (e.g., Bayes'
theorem $P(Y|X) =\frac{P(X|Y) P(Y)}{P(X)}$). While Equations
(\ref{eq:sum}) deals with probability distributions of discrete random
variables, there is also a sum rule analog for continuous random
variables and probability density functions (the syntax of the product
rule is the same):

\begin{equation*} P(X=x) = \int_Y P(X=x,\, Y=y) \, dy
\end{equation*}

where $\int_{Y}$ represents an integral over the domain of $Y$ (e.g.,
$\int_{-\infty}^{\infty}$ if $Y$ is a Gaussian random variable).

\end{document}